\documentclass[aps,letterpaper,twocolumn,prl,footinbib,floatfix,showpacs]{revtex4}
\usepackage{amsmath}
\usepackage{amsfonts}
\usepackage{amssymb}
\usepackage[scanall]{psfrag}
\usepackage{psfrag}
\usepackage{graphicx}
\usepackage{color}

\begin{document}
\newcommand{\of}[1]{\left( #1 \right)}
\newcommand{\sqof}[1]{\left[ #1 \right]}
\newcommand{\abs}[1]{\left| #1 \right|}
\newcommand{\avg}[1]{\left< #1 \right>}
\newcommand{\cuof}[1]{\left \{ #1 \right \} }
\newcommand{\pil}{\frac{\pi}{L}}
\newcommand{\bx}{\mathbf{x}}
\newcommand{\by}{\mathbf{y}}
\newcommand{\bk}{\mathbf{k}}
\newcommand{\bp}{\mathbf{p}}
\newcommand{\bl}{\mathbf{l}}
\newcommand{\bq}{\mathbf{q}}
\newcommand{\bs}{\mathbf{s}}
\newcommand{\psibar}{\overline{\psi}}
\newcommand{\svec}{\overrightarrow{\sigma}}
\newcommand{\dvec}{\overrightarrow{\partial}}
\newcommand{\bA}{\mathbf{A}}
\newcommand{\bdelta}{\mathbf{\delta}}
\newcommand{\bK}{\mathbf{K}}
\newcommand{\bQ}{\mathbf{Q}}
\newcommand{\bG}{\mathbf{G}}
\newcommand{\bw}{\mathbf{w}}
\newcommand{\bL}{\mathbf{L}}
\newcommand{\up}{\uparrow}
\newcommand{\down}{\downarrow}
\newcommand{\dl}{{\rm d}\lambda}
\newcommand{\dt}{{\rm d}t}
\newcommand{\nueff}{\nu_{\rm eff}}
\newcommand{\Nimp}{N_{\rm imp}}
\newcommand{\NLLL}{N_{\rm LLL}}

\author{Eliot Kapit}
\email[Contact:]{ek359@cornell.edu}
\author{Paul Ginsparg}
\author{Erich Mueller}
\affiliation{Laboratory of Atomic and Solid State Physics, Cornell University}
\title{Non-Abelian Braiding of Lattice Bosons}
\date{\today}

\begin{abstract}
We report on a numerical experiment in which we use time-dependent potentials to braid non-abelian quasiparticles.
We consider
lattice bosons in a uniform magnetic field within the fractional quantum Hall regime, 
where $\nu$, the ratio of particles to flux quanta, is near 1/2, 1 or 3/2. 
 We introduce time-dependent potentials which move quasiparticle excitations around one another, 
 explicitly simulating
a braiding operation which could implement part of a gate in a quantum computation.
We find that different braids do not commute for $\nu$ near $1$ and $3/2$, with Berry matrices respectively consistent with Ising and Fibonacci anyons. Near $\nu=1/2$, the braids commute.
\end{abstract}



\maketitle


\section{Introduction}

When two identical quantum mechanical particles exchange places, the wavefunction typically acquires a phase: $\theta=0$ for bosons, and $\theta=\pi$ for fermions.  Remarkably, there exist 2d systems \cite{nayaksimon,laughlinoriginal,haldaneexact,girvin,yoshoika,wenniu,wendagotto}
whose ``anyon" excitations display {\em fractional statistics}, with $\theta\neq0,\pi$.  Even more remarkably, there are models in which exchanging quasiparticles not only produces a phase, but also rotates the system between degenerate states \cite{mooreread,greiterwen,readrezayi1,readrezayi2,nayakwilczek,fradkinnayak,simonrezayi,fukane,levinwen,kitaev,kitaev2003,cooperwilkin}. Under these circumstances, exchanges may not commute. Kitaev \cite{kitaev2003} proposed using such nonabelian quasiparticles for quantum computation, with qubits constructed from the degenerate states. Quantum gates are implemented by ``braiding" the quasiparticles: using time-dependent potentials to drag the quasiparticles around one another, switching their positions. The collective nature of the encoded quantum information provides protection against various decoherence mechanisms. Here we start from a microscopic Hamiltonian, and numerically calculate the result of such a braiding experiment.  We find that even for surprisingly small systems ($4\times 4$ lattices), this procedure can be used to establish non-abelian statistics, and hence to implement quantum gates. 

Explicitly calculating the results of a braiding operation for a realistic microscopic Hamiltonian is difficult. Previous studies have focused on the properties of variational wavefunctions in the limit of all quasiparticles asymptotically far apart (separation large compared to the magnetic length) \cite{nayaksimon,wendagotto,nayakwilczek,georgiev,freedmannayak,jeongraham1,jeongraham2}. As in physical experiments, a numerical experiment must contend with finite size effects, mixing of higher bands, the location of unpinned quasiparticles, and uncertainty about both the exact many-body wavefunction and the interaction between a quasiparticle and the applied perturbation. Overcoming these difficulties is well worth the effort, since observing the braiding of two quasiparticles provides a definitive test of exchange statistics.  This numerical approach complements more indirect methods in real experiments, such as observing shot noise or interference effects in the tunneling of edge states \cite{willett}.

\section{Model}

We choose a model which is both experimentally relevant, and computationally tractable: hard-core bosons hopping on a square lattice, with phases on the hopping matrix elements corresponding to a uniform magnetic field.  This model describes Cooper pairs hopping on a Josephson junction array in a magnetic field \cite{vanderzant,fazio,ioffe} when the charging energy is large compared to the hopping energy. It also describes cold atoms in a deep optical lattice \cite{bloch} with an artificial gauge field \cite{spielman,lin,williams,cooper,liu}.  Recent developments in cold atom physics \cite{spielman} suggest that the fractional quantum Hall regime will be attained in the near future.

A general Hamiltonian for lattice bosons is
\begin{eqnarray}\label{BH}
H = - \sum_{jk} \of{ J_{jk} e^{i \phi_{jk}} a_{j}^{\dagger} a_{k} + H.C. } + \frac{U_{2}}{2} \sum_j a_j^{\dagger} a_j^{\dagger} a_j a_j \\ + \frac{U_{3}}{6} \sum_j a_j^{\dagger} a_j^{\dagger} a_j^{\dagger} a_j a_j a_j. \nonumber
\end{eqnarray}
$a_{k}^{\dagger}/a_{k}$ creates/annihilates a boson at complex coordinate $z_k$ on a square lattice with unit lattice spacing. Defining $z \equiv z_j - z_k = x + i y$ as a complex integer, $i \phi_{jk} = - \frac{\pi \phi}{2} \of{z_{j} z^{*} - z_{j}^{*} z }$ is the Peierls phase of the $\mathbf{B}$ field (with $\phi$ the density of flux quanta per plaquette).  
The
properties of this Hamiltonian depend on the form of $J_{ij}$.  The simplest model would just include nearest neighbor hopping \cite{hofstadter}.  As argued in \cite{kapitmueller}, the fractional quantum Hall states are particularly robust if we use a particular gaussian hopping,
$J_{jk} \equiv J \of{z} = J_{0} G \of{z} \exp \of{ - \frac{\pi}{2} \of{1-\phi} \abs{z}^{2} },$ where $G \of{z} = \of{-1}^{1+ x + y + x y}$ and $J_0$ is a constant. We mostly take the hard-core limit of $U_{2} \to \infty$. We define $J_{NN} = J_{0} e^{-\pi/4}$ as the energy scale of the problem. $U_{3}$ is an artificial three-body repulsion which we introduce in some calculations. We showed in \cite{kapitmueller} that the single-particle spectrum of (\ref{BH}) reproduces the continuum lowest Landau level (LLL) with $\phi L^{2}$ degenerate single particle ground states on an $L \times L$ lattice.  As explained in \cite{kapitmueller}, the longer range hoppings can be engineered by appropriately shunting the Josephson junction array, or by appropriately tailoring the optical lattice potential.  For $\phi \lesssim 1/3$ it suffices to include next-nearest-neighbor hopping. Since the lowest Landau level is preserved in (\ref{BH}), a LLL-projected calculation in the continuum whould give similar results, at the cost of more complexity in the calculation.

We add to Eq.~(\ref{BH}) a time-dependent potential $V_j(t)$, corresponding to a Hamiltonian
$
H_p=\sum_j V_j(t) a_j^\dagger a_j.
$
At time $t=0$, we take $V$ to be zero except on a few sites, where it is positive.  We slowly change $V$ such that $V_j(T)=V_j(0)$, but with two of the potential bumps  exchanged.  If quasiparticles are pinned to the defects, this will exchange them.  Experimentally, the potential $V_j$ could be engineered by gates on individual Josephson junctions, or through targeted lasers in an optical lattice. Such addressability was recently demonstrated in \cite{weitenberg}.
In our numerics we move our bumps by linearly reducing the amplitude of $V$ on one site, while linearly increasing it on a neighbor.

Under an adiabatic cyclic change of the Hamiltonian, non-degenerate states will return to themselves with an additional phase factor, while degenerate states can mix: $e^{-i H T}|\psi_i\rangle=e^{-i\int E\, \dt}\sum_j M_{ij} |\psi_j\rangle$.  Throughout we neglect the $\int E\,\dt$ term, where $E(t)$ is the instantaneous energy at time $t$.   This temporal phase can be experimentally distinguished from the geometric phase by traversing the path at different rates.  The unitary matrix $M_{ij}$ is calculated by integrating the Berry connection:
\begin{eqnarray}\label{berrycon}
M = P \exp \of{2\pi i \oint {\rm d}\lambda\, \gamma }. 
\end{eqnarray}
Here, $\gamma_{ij}= i \left < \psi_i \right | \nabla_{\lambda} \left | \psi_j \right >$ is the Berry connection matrix, the $\left| \psi_i \right>$ are a basis of degenerate states, $\lambda$ parametrizes the path, and $P$ is the path ordering symbol. While the Berry connection $\gamma$ is a gauge-dependent quantity, the matrix $M$ is physical and gauge invariant (up to joint choice of basis at the start and end points).

To numerically calculate Eq.~(\ref{berrycon}), we use a method described in \cite{pancharatnam,resta}, breaking the path into many small discrete steps, engineered to maintain the degeneracies of the spectrum.  For each point $\lambda$ on the path, we diagonalize $H$ to produce a basis $|\psi_i(\lambda)\rangle$.  This basis is not unique: the phases of $|\psi_i(\lambda)\rangle$ are arbitrary, but one can form a new basis by taking arbitrary linear superpositions of degenerate states.  We fix this arbitrariness by choosing  $\left < \psi_{i} \of{\lambda} | \psi_{j} \of{\lambda + \dl} \right > = \delta_{ij} + O \of{\dl^{2} }.$  The Berry matrix is then
\begin{equation}
M_{ij}=\langle \psi_i(\lambda_f)|\psi_j(0)\rangle\ .
\end{equation}
Following \cite{resta}, we generate the states $| \psi_i(\lambda+\dl) \rangle = |\psi_i(\lambda)\rangle=\sum_j (A^{-1})_{ij} |\tilde \psi_j(\lambda)\rangle$ by first determining the eigenstates $|\tilde \psi_i(\lambda+\dl)\rangle$ using a generic diagonalization algorithm,
and then calculating the overlap matrix $A_{ij}=\langle \psi_i(\lambda)|\tilde \psi_j(\lambda+\dl)\rangle$. Since $A$ will be unitary only up to corrections of order $\dl$, we perform a Gram-Schmidt orthogonalization at each step.

In Fig.~\ref{sqbraidpaths}, we illustrate the initial configurations of the impurities and some of paths over which we move them.  
We use relatively small systems: between 3 and 9 particles on lattices of up to 24 sites with periodic boundary conditions; with the hard core constraint the largest Hilbert spaces studied contained about 50,000 states.  While state-of-the art algorithms on high performance computers would allow us to study larger systems, we find that finite size effects are already sufficiently small on these modest grids, presumably due to the robust nature of the topological effects of interest. Our algorithm was implemented in \textit{Mathematica} on a desktop computer.   

\section{Results}

The results of our braiding calculations are summarized in table I. In all cases, the applied impurity potentials are strong. We assign each state an effective filling fraction $\nueff = N/\NLLL$, where $\NLLL$ is the number of single particle states in the LLL in the presence of the impurities. In every case studied, for $\Nimp$ impurities $\NLLL = N_{\phi}-\Nimp$ (where $N_{\phi}$ is the number of flux quanta), showing that a full quasihole (QH) is pinned at each impurity, and in the thermodynamic limit $\nueff \to \nu$. Each quasihole is a first order zero of the many-body wavefunction and binds a single flux quantum. These full QHs will be supplemented by a number of non-abelian fractional quasiholes at the appropriate filling fractions. In the table, each unitary braid matrix $M$ is denoted by a pair of phases ($p_{1},p_{2}$), where $e^{i \pi p_{1} }$ and $e^{i \pi p_{2}}$ are the eigenvalues of $M$. For cases with more than 2 impurities, we label the exchange of impurities $i$ and $j$ (as labeled in Fig.~1) by $R_{ij}$.

The simplest case $\nueff=1/2$ provides an excellent test of the algorithm, since we know (in the absence of a perturbing potential) that both the ground state wavefunction, and its quasihole excitations, are given exactly by Laughlin's variational ansatz  \cite{kapitmueller}.  On the torus the ground state is twofold degenerate \cite{oshikawa}.  
Excitations about these two degenerate ground states states require overcoming an energy gap $\Delta \sim J_{NN}$. The quasiholes are abelian anyons, and the Berry matrix in the ground state subspace should be the identity times a phase of $\pm\pi/2$, depending on the direction of the exchange path \cite{wenniu,wendagotto}. 
This is consistent with our numerical studies of the path in Fig.~1(a).
Since a complete braid of one quasihole around another is equivalent to two exchanges, we find a phase of $\pi$ for the path 1(c).  
As expected, when we introduce more impurities, we find that near $\nu_{\rm eff}=1/2$ all braids commute. 

A generic potential splits the two-fold degeneracy of the ground state by a small energy $\epsilon$. We attribute these splittings to interactions between the quasiparticles when they are moved close to one another.  By optimizing the shapes of the potential at each time step, we can make $\epsilon<0.02\Delta$ for all points in the $\nueff = 1/2$ braid. While largely irrelevant for $\nueff=1/2$, this optimization can be crucial for producing sensible results near $\nueff=1$ or 3/2.  If the trajectory is traversed in a time $T$ such that $\hbar/\Delta \ll T \ll \hbar/\epsilon$ these splittings have no physical effect.  We therefore neglect them when calculating $M$.  We expect that the splittings can be further reduced by using larger systems.  Detailed graphs of our optimized potentials are shown in the supplemental information for this paper \cite{supplemental}.

\begin{figure}
\includegraphics[width=3.25in]{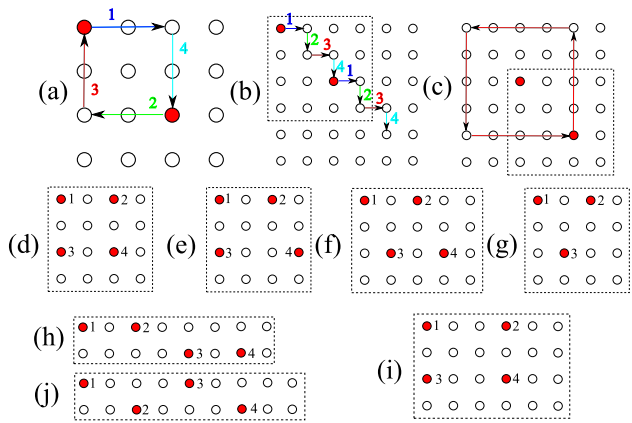}
\caption{(Color online) (a)--(c) Exchange paths used to braid quasiparticles on various lattices. In each path, the impurities (shaded red) are incrementally moved along the segments (1,2,3...) until they return to their starting positions, exchanged. The dashed box represents the periodic lattice boundary. (d)--(i) show the initial configurations of the impurities for the 3- and 4- impurity braids.}\label{sqbraidpaths}
\vglue-10pt
\end{figure}

\begin{table}
\hskip-10pt\vbox{
\begin{tabular}{| c | c | c | c | c | c | c |}
\hline
Lattice & $N$ & $N_{\phi}$ & $\Nimp$ & GFS & $\nueff$ & Braid Path/Phases (all $\times \pi$) \\

\hline
\multicolumn{7}{|l|}
{\em Abelian}\\
\hline
$4\times 4$ & 3 & 8 & 2 & G & 1/2 & (a) (0.49,0.49), (c) (0.99,0.99) \\
$4\times 4$ & 6 & 8 & 2 & F & 1 &(a) (0,0.99), (b) (0,1) \\
$4 \times 4$ & 7 & 8 & 2 & G & 7/6 & (b) (0,1) \\
\hline
\multicolumn{7}{|l|}
{\em Non-abelian}\\
\hline
$6 \times 4$ & 4 & 8 & 4(i) & S & 1 & $R_{12},R_{34}:$ (0.28,-0.28) \\ 
& & & & & & $R_{24}: (-0.26,-0.75)$\\
& & & & & & $R_{13}: (0.22,-0.22)$\\
$4 \times 4$ & 7 & 8 & 3(g) & F & 7/5 & $R_{13},R_{23}$: (0.08,0.73) \\
& & & & & & $R_{12}:$ (0.08(1),0.76(4))\\
\hline
\multicolumn{7}{|l|}
{\em Ambiguous}\\
\hline
$4\times 4^{*}$ & 4 & 8 & 4(d) & F & 1 & $R_{12},R_{13},R_{24},R_{34}:$ (0,1) \\

$4 \times 4$ & 7 & 10 & 4(e) & F & 7/6 &  $R_{12},R_{13},R_{24},R_{34}:$ \\ 
& & & & & & (0.25(2),-0.25(2)) \\
$5 \times 4$ & 4 & 10 & 4(f) & G & 2/3 & $R_{12},R_{34}:$ (-0.75,0.75),  \\
$8 \times 2$ & 6 & 8 & 4(h) & F & 3/2 & $R_{12},R_{34}:$ (0.32,-0.32)\\
& & & & & & $R_{23},R_{14}: (0,1)$\\
$9 \times 2$ & 9 & 10 & 4(j) & F & 3/2 &  $R_{34}:$ (0.69,-0.69) \\
\hline
\end{tabular}}
\caption{The results of our numerical braiding studies. Here, $N$ is the total particle number, $N_\phi$ is the total number of flux quanta, and $\Nimp$ impurity sites have a repulsive potential applied. ``GFS" refers to whether the degenerate pair of eigenstates are the ground (G), first excited (F) or second excited (S) states. The braids are each characterized by a unitary matrix with eigenvalues $e^{i \pi p_{1} }, e^{i \pi p_{2}} \to \of{ p_1, p_2}$. The exchange paths are shown in Fig.~1, with $R_{ij}$ denoting the exchange of impurities $i$ and $j$. The algebras in the non-abelian cases approximate those described in the text \cite{georgiev,hormozi}; cases labeled as ambiguous contain non-commuting paths but the transformations associated with these paths depended on the details of the path and/or did not match the analytical predictions. Due to finite size splitting, not all paths were accessible on all lattices; only paths which led to a sensible braid and which were stable against small changes in the impurity strength $V_j$ are quoted here. Hard core interactions $\of{ U_2 = \infty}$ were used in all cases except $4\times4^{*}$, where we also used $\of{U_2 = 0, U_3 = \infty}$. These two interactions gave nearly identical results.
}\label{braidtab}
\vskip-14pt
\end{table}

The physics near $\nueff = 1$ and 3/2 is richer. At $\nueff=1$ for $U_{2}$ small, all particles are in the lowest Landau level and the ground state $\Psi_{\rm G}$ has a large overlap \cite{cooperwilkin} with the Moore-Read (M-R) Pfaffian state $\Psi_{\rm MR}$ \cite{nayaksimon,mooreread,greiterwen,readrezayi1,nayakwilczek,fradkinnayak}, a state with non-Abelian excitations. We typically perform our calculations using hard-core interactions, where mixing with excited bands is significant and the overlap is smaller: $\abs{\langle \Psi_{\rm MR} | \Psi_{\rm G} \rangle} < 0.3$. Despite the small overlaps, one expects that the ground state with hard-core interactions is adiabatically connected to the M-R state and should share topological invariants such as exchange statistics. The M-R state is the exact ground state of a Hamiltonian with repulsive three body interactions ($U_{2} = 0$, $U_{3} > 0$). As we expand on below, we find excellent agreement between calculations using the two and three body interactions.

The M-R state is gapped and has two types of fundamental vortex excitations. In addition to the full QHs described earlier, the M-R state has half-quasihole (HQH) excitations, which bind half a flux quantum, partially exclude particles from their location, and are non-abelian Ising anyons \cite{nayaksimon}. Wavefunctions of the M-R type with $2n$ HQHs are $2^{n-1}$-fold degenerate \cite{nayakwilczek} in the limit that all the HQHs are far apart. Given that we use strong impurity potentials ($V_j \geq J_{NN}$) we expect each repulsive impurity will bind a full QH and a half quantum vortex. Exchanging two HQHs performs a $\pi/2$ rotation within the degenerate subspace, and the rotations produced by exchanging different pairs of HQHs do not generally commute. In particular, for four HQHs, it was shown \cite{bravyi,georgiev} that in the appropriate basis and ignoring Abelian phases, the braids can be written as
\begin{eqnarray}\label{nu1}
R_{12} = R_{34} = e^{-i \frac{\pi}{4} \sigma_y }, \; R_{13} = R_{24} = e^{-i \frac{\pi}{4} \sigma_x }.
\end{eqnarray}

To estimate the overlap of the unitary transformations which result from our braids with the predictions of the analytical theories of Bose quantum Hall states, we use the matrix overlap measure $\of{M_1,M_2} \equiv |{\rm tr}(M_1M^\dagger_2)|/2$.  
This quantity is insensitive to overall phases, and we consider two unitary matrices to be equivalent if $| {\rm tr}(M_1M^\dagger_2) |/2 = 1$.

For the case of $N = 4, N_{\phi} = 8$ and $\Nimp = 4$ on the $6 \times 4$ lattice (where two impurities need never be nearest or next-nearest neighbors in a braid), our numerical results are in remarkable agreement with eq. (\ref{nu1}). Labeling the analytical predictions by $R$ and the numerical matrices $M$, we have $\of{R_{12},M_{12}} = \of{R_{34},M_{34} } = 0.99$,  $\of{R_{24}, M_{24} } = 0.98$ and $\of{R_{13},M_{13}} = 0.97$. 

When impurities are allowed to approach more closely, however, the numerical results diverge from the analytical predictions, and in many cases, the exchange of two strong impurities produces a rotation by $\pi$. We conjecture that this represents the exchange of two pairs of HQHs, which either do not sit directly on the impurities but move with them as they are exchanged, or experience tunneling events when impurities move too close to one another. For the case of a $4\times 4$ lattice with $N=4$, $N_{\phi}=8$ and $\Nimp=4$, we obtained identical results when considering the ordinary hard-core two-body or a hard core three-body interaction, where the M-R state is the exact ground state. For $N=7$ and $N_{\phi}=10$ on the same lattice (Fig.~\ref{sqbraidpaths}e), we consistently obtained rotations by $\of{0.5 \pm 0.03}\pi$, but the matrices which resulted were not straightforwardly related to the analytical predictions in eq. (\ref{nu1}), and depended strongly on the path by which a pair of impurities were exchanged. These results show that the precise relationship of the non-abelian vorticies to the impurities is subtle \cite{toke,huwan,prodan,storni}. Further, they reveal that the Berry matrices can be strongly modified for paths where impurities come close together. Surprisingly, the degeneracies are not necessarily broken by these close approaches.

Finally, near $\nueff = 3/2$ (fig.~\ref{sqbraidpaths}g,h,j), we obtained a result consistent with the predictions for a Fibonacci anyon theory \cite{hormozi}, the effective theory of the Read-Rezayi state at $k=3$ \cite{readrezayi2}. Previous numerical studies of continuum bosons in the LLL \cite{cooperwilkin} have found strong evidence for this state, a particularly exciting result since Fibonacci anyons are capable of universal topological quantum computing.
Comparing our numerically derived matrices at ($N = 7$, $N_{\phi} = 8$ and $\Nimp = 3$) with the results of Hormozi et al.\ \cite{hormozi}, we obtained $ \of{R_{13}, M_{13} }= 0.99$ and $\of{R_{23},M_{23} } = 0.90$. However, for the exchange of impurities 1 and 2, we found two sensible paths (a) and (b); in path (a) impurity 3 was allowed to move during the braid and in (b) it was not. We found that $\of{R_{12},M_{12}(a)}=0.93$, but $\of{R_{12},M_{12}(b)}=0.69$ and $\of{M_{12}(a),M_{12}(b)}=0.46$. We conjecture that this disagreement was due to tunneling events when the impurities were only next-nearest neighbors. For the $8 \times 2$ and $9 \times 2$ lattices, we obtained rotations of nearly $3\pi/5$ as predicted, but the resulting matrices had little overlap with those predicted from the Fibonacci anyon theory.


\section{Summary and Conclusions}

In summary, we have numerically studied a realistic model, eq. (\ref{BH}), which has anyon excitations at filling fraction $\nueff = 1/2$, and non-abelian anyons at $\nueff = 1$ and 3/2 analogous to those in the Moore-Read and Read-Rezayi states. These results suggest adiabatic continuity between the states of our lattice model with hard-core interactions and those found purely in the LLL \cite{cooperwilkin}, to which our model reduces in the limit of weaker on-site interaction. We have also shown that surprisingly small lattices can reproduce infinite-system predictions, without resorting to trial wavefunctions. This robustness is likely related to the topologically protected nature of the states, and is encouraging for future experiments.

The most intriguing implication of our result is in quantum computation. In recent years, a wealth of theory \cite{nayaksimon,bravyi,bondersonkitaev,freedmannayak,georgiev} has shown that the M-R state of electrons at $\nu = 5/2$ could be used to construct topologically protected quantum memory and quantum computing operations, and has described potential implementations. While non-abelian statistics in the $\nu = 5/2$ state have not yet been confirmed experimentally, the fact that the $\nu = 1$ M-R state and the $\nu = 5/2$ M-R state are in the same universality class implies that the theory for manipulating quasiholes in the $\nu = 5/2$ electron gas can be applied directly to our lattice boson system. Our $\nu=3/2$ results are even more exciting since the Read-Rezayi states can be used to construct a universal quantum gate set. Implementing our model in a Josephson junction array would open a new area of physics to study topological noise protection and non-abelian statistics, since for $\phi \leq 1/4$ three non-abelian plateaux ($\nu=1,3/2,$ and $2$) could be studied in the same experiment. The ability to individually address any lattice site would provide an unprecedented ability to manipulate quasiholes \cite{freedmannayak}, potentially creating a truly universal ``quantum loom."

\section{Acknowledgments}

We thank Chetan Nayak, Andrei Bernevig, Chris Laumann and Chris Henley for useful discussions. This work was supported by an Army Research Office grant with funding from the DARPA OLE program, by NSF grant PHY-1068165 and by the Department of Defense (DoD) through the National Defense Science and Engineering Graduate (NDSEG) program.

\end{document}